\documentclass[fleqn]{article}
\usepackage[hyphenbreaks]{breakurl}
\usepackage[hyphens]{url}
\usepackage{mathtools}
\usepackage{bm}
\usepackage{cancel}
\usepackage{amsmath}
\usepackage{tcolorbox}
\usepackage[font={small,it}]{caption}

\begin{document}

	\title{A Systematic Method for On-The-Fly Denormalization of Relational Databases}
	\author{Sareen Shah\\Zucker School of Medicine at Hofstra/Northwell}
	\date{}
	\maketitle
	\renewcommand{\abstractname}{\vspace{-\baselineskip}}

	\begin{abstract}
	    \noindent \textbf{Abstract --} Normalized relational databases are a common method for storing data, but pulling out usable denormalized data for consumption generally requires either direct access to the source data or creation of an appropriate view or table by a database administrator. End-users are thus limited in their ability to explore and use data that is stored in this manner. Presented here is a method for performing automated denormalization of relational databases at run-time, without requiring access to source data or ongoing intervention by a database administrator. Furthermore, this method does not require a restructure of the database itself and so it can be flexibly applied as a layer on top of already existing databases.
	\end{abstract}
	
	\section*{Introduction}
	
	Normalized relational databases took their form initially from a need to reduce duplication of records and to organize data in a manner whereby updating or adding records would not lead to data inconsistencies \cite{ibm}. However, this structure can lead to difficulty with the actual retrieval of usable information. For wide databases, it can be an arduous process to discover the logical sequence of tables that must be joined to bring information from different ends of the database together, especially when many combinations of different data elements are desired. A more important limitation is that an end-user of the data without direct access is reliant upon the database administrator creating the denormalized tables \textit{a priori} for any data exploration/analysis desired.
	
Relational databases remain the most common method of storing data \cite{dbengines}. Data collected for research purposes are also usually stored/provided using a relational schema. Thus, there is a role for creating a technique for on-the-fly denormalization, to allow for maximal flexibility for information retrieval by end-users and to unburden the database administrator. The algorithms provided here will enable an end-user to be able to combine any number of columns of interest from any number of tables, and automatically construct the join sequences required to perform this denormalization \textit{ad hoc}.

	\section*{Illustrative Example}
	
	We will use a modified version of the Northwind Traders database to demonstrate scenarios where an on-the-fly denormalization algorithm can be useful. The Northwind Traders database was originally created by the Microsoft Corporation, and the modified files can be found at https:\slash\slash github.com\slash sareeneng\slash NorthwindModified. The modified dataset consists of 11 tables that are organized in a typical normalized fashion, although some of the relationships have been constructed in a way to highlight certain points in the discussion to follow, and so it is not a fully optimized data representation. For simplicity, only the joining columns and a few other columns are shown in \textbf{Figure 1}.
	
	\begin{figure}[ht]
		\centering
		\includegraphics[width=1\linewidth]{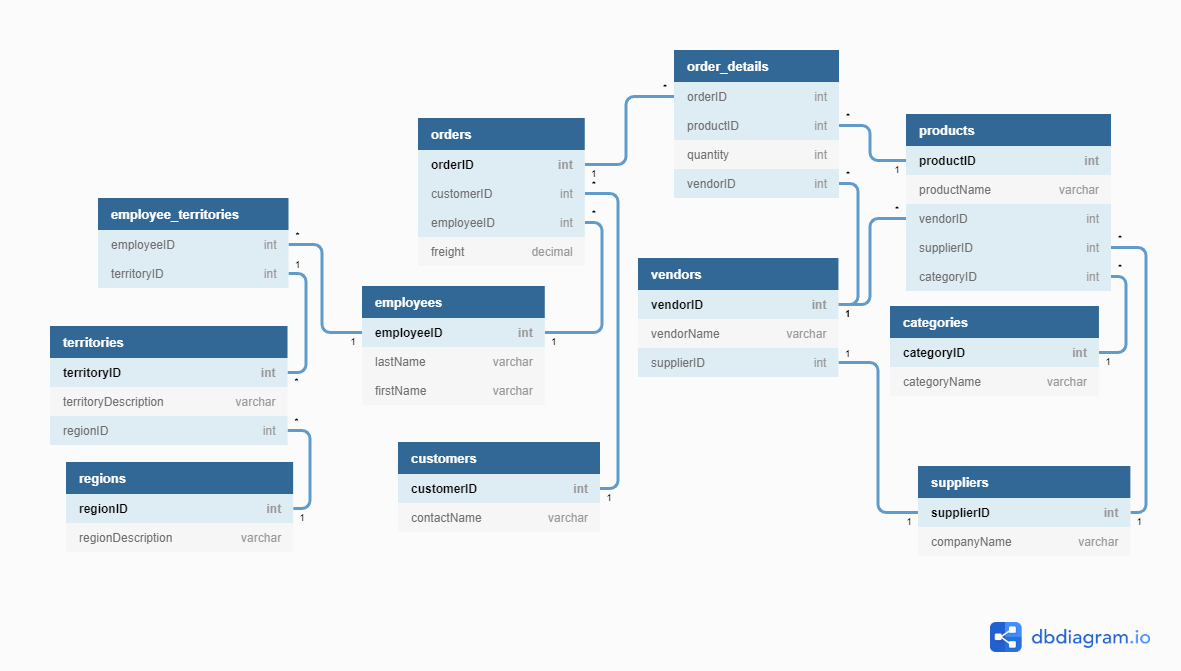}
		\caption{Schema of the modified Northwind dataset. *s represent many-to-one relationships.}
		\label{fig:northwind-compact}
	\end{figure}
	
	Suppose an end-user is interested in finding which employees and customers have interacted with each other. From the diagram, it is clear that the way to accomplish this is to start with the \textit{orders} table, and then join the \textit{employees} and \textit{customers} tables in. A database administrator might create a view to bring the data from these two tables together if this is a frequently requested combination of variables. This requires anticipation on the part of the administrator.
	
	If the end-user wants to explore the database and create many different combinations of variables, then these must also be anticipated. For example, the user might want to know what territories the orders are originating from, or see which vendors served which customers. The user may also want many more variables pulled in together, such as quantity and names of the products sold by each employee. Notably, some of these combinations may not be possible; orders are linked to employees, but employees may serve multiple territories so there is no way to definitively link the tables \textit{territories} to \textit{orders}.
	
	One (unfeasible) option is to anticipate all potential combinations that a user may be interested in and create the associated denormalized tables (e.g. with a star schema). This becomes an exponentially laborious task the wider the database and the more variables that a user might want to pull together, and the vast majority of these cases will be extraneous. The on-the-fly denormalization algorithm presented here avoids the need to anticipate end-user data requests.
	
	\section*{Graph Structure}
	
	A simple way to represent the relational database is by using a directed graph structure that will then allow for the path-finding required. Tables are represented as the nodes, and foreign key links are represented as the edges between nodes. The box details the steps for constructing the graph, and \textbf{Figure 2} shows the resulting graph for the modified Northwind dataset.

	\begin{tcolorbox}
		\textbf{Directed graph representation of relational database}
		
		\noindent For each foreign key \textit{FK} link, use the following rules to construct the graph $G$ where the nodes are tables and the edges are the foreign key links. Let a \textit{1-1} relationship mean that both tables have only unique values in their respective joining columns, let a \textit{M-1} relationship mean that the first joining column has non-unique values and the second joining column has only unique values, and let a \textit{M-M} relationship mean that both joining columns have  non-unique values.
		\begin{itemize}
			\item If node $A$ is already connected to $B$ along a different $FK$, see text.
			\item If $A.FK_{A/B}$ has a \textit{M-M} relationship with $B.FK_{B/A}$\begin{itemize}\item Do not connect node $A$ to node $B$\end{itemize}
			
			\item If $A.FK_{A/B}$ has a \textit{M-1} relationship with $B.FK_{B/A}$\begin{itemize}\item $A\xrightarrow{FK} B$ \end{itemize}
			\item If $A.FK_{A/B}$ has a \textit{1-1} relationship with $B.FK_{B/A}$\begin{itemize}\item $A \xleftrightarrow{\text{FK}} B$ \end{itemize}
		\end{itemize}
	\end{tcolorbox}
	
	\begin{figure}[ht]
		\centering
		\includegraphics[width=1\linewidth]{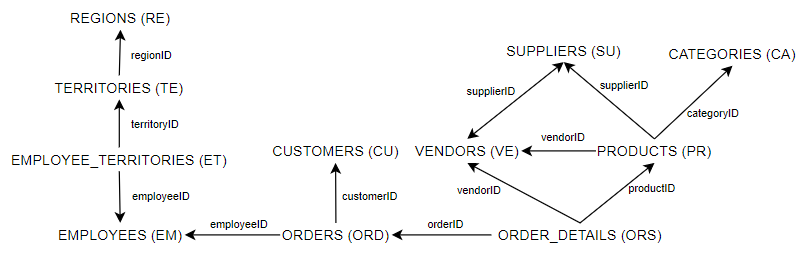}
		\caption{Directed graph representation of the modified Northwind dataset. Many-to-one relationships are connected in one direction, and one-to-one relationships are connected in both directions.}
		\label{fig:northwindgraph}
	\end{figure}
	
	The columns used to join tables together can be classified as either ``many'' (non-unique values) or ``one'' (only unique values). The method for determining whether the columns have only unique values will depend on the database management system. The links between tables can then be manually specified, and the directionality of the arrows determined based on the columns classification in both tables. In the case where there are multiple foreign keys between two tables, linking options will depend on the use-case. A database administrator may decide that only one of the foreign keys is relevant, or all links may be created and an option may be provided to the user at run-time as to which specific connection to use for the query.
	
	There are a few ways to minimize the amount of manual labor involved in creating these links, including:
	\begin{enumerate}
    	\item If the database is stored in a typical database management system with foreign keys already set up, then these foreign keys can be used to create the connections.
		\item Many databases are structured such that column names have a consistent meaning across tables. In this case, one can iterate over all column names and whenever a column is found in multiple tables, then a link can be created with directionality based on the classification described above if the uniqueness of the columns is known and will not change.
	\end{enumerate}

	Once the graph has been created, then the connections can be stored and the graph regenerated from this metadata at any time. At this point, two path-finding algorithms need to be implemented:
	\begin{enumerate}
    	\item \textbf{Find if a path exists at all between any two nodes.}
	    \item \textbf{Find all simple paths between any two nodes} - a simple path is one that connects the source to the target without traversing any interceding nodes more than once.
	\end{enumerate}
	
	Many programming languages have packages/modules that can construct these graphs and implement the path-finding algorithms needed. For example, Python users can install the \textbf{networkx} package \cite{networkx}.

	\section*{Denormalization Algorithm}
	The basic algorithm builds on the idea that one needs to find an origin node from which a simple path can be drawn to all individual desired nodes. The origin node may also be one of the desired nodes. For example, if the user is interested in data from the \textit{suppliers} and the \textit{categories} tables, a path cannot be drawn directly from one to the other. However, the origin node \textit{products} has a path to each individual desired table, and so it can serve as the left-most table in the join sequence. There is the potential for a significant amount of redundancy; \textit{order\_details} could have served as an origin node, but it does not add more useful information and adds more complexity if the user is only interested in \textit{suppliers} and \textit{categories}, while \textit{products} is a necessary table in the join sequence. The steps below describe how to eliminate the redundancy.
	
	\subsection*{Find paths between two given tables}
	
	The first building block is to find all simple paths between an origin node and each desired table. Some of the paths found may include extraneous tables (which may reduce useful information while adding to complexity), and so the redundancy should be removed (\textbf{Algorithm 1}). First, sort the list of paths found from shortest to longest, then eliminate all paths that are a superset of any previous path. For example, the simple paths from \textit{order\_details} to \textit{suppliers} are $ORS\rightarrow VE\rightarrow SU$, $ORS\rightarrow PR\rightarrow SU$, and $ORS\rightarrow PR\rightarrow VE\rightarrow SU$. The first two paths are kept because neither is a superset of the other, and it is unknown at this time which is the optimal path (e.g. going through \textit{vendors} might yield more rows or more relevant data than going through \textit{products}, or vice-versa). However the last path can be removed because there is no possibility of it being more optimal than either of the first two paths. 
	
	Note that this is not the same as finding edge disjoint paths. If one path is $A\rightarrow B\rightarrow C\rightarrow E$ and another is $A\rightarrow B\rightarrow D\rightarrow E$, these paths share an edge but it is not clear if one is inferior to the other and so both should be kept.
	
	\begin{tcolorbox}
		\textbf{Algorithm 1: Finding reduced simple paths between two specific nodes}
		\begin{enumerate}
			\item Let $L$ be a list of ordered pairs representing the edges of a single simple path $T_s\rightarrow T_x \rightarrow T_y \rightarrow...\rightarrow T_w \rightarrow T_z \rightarrow T_e$ as such:\\ $L = [(T_s, T_x), (T_x, T_y), ..., (T_w, T_z), (T_z, T_e)]$
			\item Let $M(T_s, T_e)$ be a list of all $L$ from $T_s$ to $T_e$
			\item Sort $M$ by the number of unique nodes in $M_i$ in ascending order
			\item Remove all $M_i$ from $M$ where the nodes in $M_i$ are a superset of the nodes in any prior $M_{1...i-1}$ for all $i=2...n$
		\end{enumerate}
	\end{tcolorbox}
    \vspace{-5mm}
	\subsection*{Finding join sequences between multiple tables}
	
	If an origin node exists from which a simple path can be drawn to all desired nodes, then at least one join sequence can be generated. The basic technique is to first take all the paths between all possible origin nodes and each of the desired tables. The origin node might be one of the desired nodes. Then calculate the cartesian product to generate all possible valid combinations of simple paths, and in a similar fashion as above, sort the candidate join sequences by the number of unique tables to remove those that are simply supersets of prior sequences. These steps are outlined in \textbf{Algorithm 2}.
	
	The operation of finding all simple paths can be computationally expensive for very large graphs. A significant optimization can be done by first checking to see if a path exists at all from the candidate origin node to each of the desired nodes using Dijkstra's algorithm, and if not, then move on to the next candidate.
	
	The result will be a list of lists, with each individual list comprised of a set of ordered pairs representing a join sequence. For example: $[(T_1.C_1, T_2.C_2)$, $(T_1.C_3, T_3.C_4), ...]$ represents ``\small{$T_1$ JOIN $T_2$ ON $T_1.C_1=T_2.C_2$ JOIN $T3$ ON $T1.C3=T3.C4$ JOIN ...}'' . For brevity, the above can also be represented as $[(T_1, T_2), (T_1, T_3)]$
	
	A fully worked example is provided in the Appendix, where a join sequence is found between \textit{customers}, \textit{suppliers}, and \textit{categories}. Note that this is different from the examples provided above in order to specifically highlight all the steps of the algorithm. The join sequence generated by this algorithm uses the \textit{order\_details} table as the origin/left-most table. Of note, the algorithm eliminates the potential join sequence that makes use of the path from $ORS\rightarrow VE\rightarrow SU$ because in order to capture \textit{categories}, the \textit{products} table must be included and since \textit{suppliers} is directly related to \textit{products}, there is no need to use \textit{vendors}. If \textit{categories} was not required, then two valid join sequences would have been produced.
	
	\begin{tcolorbox}
		\textbf{Algorithm 2: Finding join sequences between multiple tables}
		\begin{enumerate}
			\item Let the list of tables desired be represented as $D = [T_1, ..., T_n]$
			\item Let $H$ be an empty list of valid join sequences
			\item Let the list of all tables in the graph be represented as $S = [S_1, ..., S_n]$. For all $S_{i...n}$: let $V_{i,1} = M(S_i, D_1), V_{i,2} = M(S_i, D_2), ..., V_{i,n} = M(S_i, D_n)$ where $S_i \neq D_j$. If $V_{i,j}$ is not empty for all $j = 1...n$, then:
			\begin{enumerate}
				\item Let $P_i$ = $\{V_{i,1}\} \times \{V_{i,2}\} \times ... \times \{V_{i,n}\}$
				\item For each path combination $P_{i,k}$:
				\begin{enumerate}
					\item Flatten $P_{i,k}$ to a single list of ordered pairs.
					\item Remove duplicated ordered pairs from $P_{i,k}$.
					\item Add $P_{i,k}$ to $H$
				\end{enumerate}
			\end{enumerate}

			\item Sort $H$ by the number of unique tables traversed by $H_i$
			\item Remove all $H_i$ from $H$ where the tables traversed in $H_i$ are a superset of the tables in any prior $H_{1...i-1}$ for all $i=2...n$
		\end{enumerate}
	\end{tcolorbox}

	\section*{Discussion}
	
	The algorithm presented here allows for end-user creation and consumption of denormalized data without requiring a restructuring of the normalized database and without requiring access to the source data. Once the initial set-up has been completed by a database administrator, data can be pulled from different tables with the appropriate joins created automatically through the steps detailed above. Furthermore, this algorithm is relatively simple to implement in various programming languages. A plausible use-case scenario could be a dashboard where an end-user can select data elements of interest and choose filters, and then the resulting denormalized data can be used to construct visualizations. 
	
	There are several limitations, some of which have been described above. 
	\begin{itemize}
		\item The links need to be set up manually, but this is a one-time task. Ideally this should be done by an administrator familiar with the database structure (especially if the database is live) as this person can identify which are the important links between tables, and classify the relationships appropriately based on defined constraints and foreign keys. If the data is provided as static files and there is no database administrator (e.g. research data provided for distribution in \textit{.csv} format), then the column data itself can be analyzed to determine if there are unique/non-unique values present, and then the appropriate relationships can be set up.
		
		\item For very large and highly connected graphs, the path-finding may be computationally expensive. Some options to ameliorate this include storing the paths when found to avoid having to recalculate them, ensuring a path exists using Djikstra's algorithm, and setting a maximum search depth. Once the paths have been found for a pair of nodes, they can be stored for future use so that they do not need to be recalculated every time.
		
		\item There is no specific support for self-referencing tables, though the algorithm could be modified to incorporate this.
		
		\item In the case where there are multiple valid join sequences, a decision needs to be made on which data to provide. Options are to 1) prioritize paths that traverse mandatory over optional foreign keys, 2) provide the ``union distinct'' output of all possible results, 3) provide each result separately, or 4) provide the table that has the most records in it.
	\end{itemize}

\section*{Acknowledgments}
I would like to thank Dr. Alysia Flynn for her thorough review and analysis of the algorithm, as well as Dr. Nelson Sanchez-Pinto for his guidance and mentorship.

\bibliographystyle{ieeetr}
\bibliography{biblio}

\newpage
\section*{Appendix}

\subsection*{Worked Example}

\textbf{Find a join sequence between \textit{customers}, \textit{suppliers}, and \textit{categories} (from Figure 1)}

\begin{enumerate}
	\item $D$ = $[CU, SU, CA]$
	\item $H$ = $[\;]$
	\item $S = [ORD, ORS, TE, ET, EM, RE, CU, PA, VE, SU, CA]$\\
	\rule{2cm}{0.4pt}\\
	$V_{1,1} = M(S_1, D_1) = M(ORD, CU)$
	\begin{itemize}
		\item $L_1 = [(ORD, CU)]$, this is the only simple path.
		\item $M(ORD, CU) \rightarrow [\;[(ORD, CU)]\;]$
	\end{itemize}
	$V_{1,2} = M(S_1, D_2) = M(ORD, SU) \rightarrow$ no simple paths found. Stop and move onto the next candidate origin node.\\
	\rule{2cm}{0.4pt}\\
	$V_{2,1} = M(S_2, D_1) = M(ORS, CU)$
	\begin{itemize}
		\item $L_1 = [\;(ORS, ORD), (ORD, CU)\;]$
		\item $V_{2,1} = M(ORS, CU)\rightarrow [\;[(ORS, ORD), (ORD, CU)]\;]$
	\end{itemize}
	$V_{2,2} = M(S_2, D_2) = M(ORS, SU)$
	\begin{itemize}
		\item $L_1 = [(ORS, PR), (PR, SU)]$
		\item $L_2 = [(ORS, VE), (VE, SU)]$
		\item $L_3 = [(ORS, PR), (PR, VE), (VE, SU)]\rightarrow$ nodes set is $\{ORS, PR, VE, SU\}$ which is a superset of the nodes in $L_1$ so eliminate it. Note that it is also a superset of $L_2$ but the first condition is enough to eliminate.
		\item $V_{2,2} = M(ORS, SU)\rightarrow \begin{aligned}[t] [\;&[(ORS, PR), (PR, SU)]\\&[(ORS, VE), (VE, SU)]\;]\end{aligned}$
	\end{itemize}
	$V_{2,3} = M(S_2, D_3) = M(ORS, CA)$
	\begin{itemize}
		\item $L_1 = [(ORS, PR), (PR, CA)]$
		\item $V_{2,3} = M(ORS, CA) \rightarrow [\;[(ORS, PR), (PR, CA)]\;]$
	\end{itemize}
	\begin{align*}
	P_2  &= \{V_{2,1}\} \times \{V_{2,2}\} \times \{V_{2,3}\} \\
	&= \begin{aligned}[t]
	[\;&[\;[(ORS, ORD), (ORD, CU)],\;[(ORS, PR), (PR, SU)],\;[(ORS, PR), (PR, CA)]\;]\\
	&[\;[(ORS, ORD), (ORD, CU)]\;, [(ORS, VE), (VE, SU)],\;[(ORS, PR), (PR, CA)]\;]\;]
	\end{aligned}
	\\
	&\xrightarrow{flatten}
	\begin{aligned}[t]
	[\; &[(ORS, ORD), (ORD, CU), (ORS, PR), (PR, SU), \cancel{(ORS, PR)}, (PR, CA)]\\
	&[(ORS, ORD), (ORD, CU), (ORS, VE), (VE, SU), (ORS, PR), (PR, CA)]\;]
	\end{aligned}
	\end{align*}
	\textbf{Add $P_{2,1}$ and $P_{2,2}$ to $H$}\\
	\rule{2cm}{0.4pt}\\
	$V_{3...11}$ have at least one $M(S_i,D_j) = [\;]$, so no new valid join sequences are found in this case, and $H$ is fully populated from $ORS$ as the origin node.
	\item $\begin{aligned}[t]
		H \xrightarrow{sorted} [\; &[(ORS, ORD), (ORD, CU), (ORS, PR), (PR, SU), (PR, CA)]\\
		&[(ORS, ORD), (ORD, CU), (ORS, VE), (VE, SU), (ORS, PR), (PR, CA)]\;]\end{aligned}$\\
		$H_1$ has 6 unique nodes $\{ORS, ORD, CU, PR, SU, CA\}$\\
		$H_2$ has 7 unique nodes $\{ORS, ORD, CU, PR, SU, CA, VE\}$
		
	\item The nodes in $H_2$ are a superset of the nodes in $H_1$ and so it can be eliminated. This leave only one join sequence:\\\\ $[(ORS, ORD), (ORD, CU), (ORS, PR), (PR, SU), (PR, CA)]$\\\\
	which can be translated as follows:\\\\
	\texttt{$ORS$ JOIN $ORD$ ON $ORS.orderID = ORD.orderID$ JOIN $CU$ ON $ORD.customerID = CU.customerID$ JOIN $PR$ ON $ORS.productID = PR.productID$ JOIN $SU$ ON $PR.supplierID = SU.supplierID$ JOIN $CA$ ON $PR.categoryID = CA.categoryID$}
	
\end{enumerate}
\end{document}